\documentclass[floatfix,twocolumn,aps,prd,showpacs,10pt,superscriptaddress]{revtex4-1}
\usepackage{amsfonts}
\usepackage{epsfig}
\usepackage{graphicx}
\usepackage[latin1]{inputenc}
\usepackage{float}
\usepackage{color}
\usepackage{amsmath,amssymb}
\usepackage{natbib}
\begin{document}
\title{Casimir effect via a generalized Matsubara formalism}
\author{Andreson L.C. Rego}
\email{andresonlcr@gmail.br}
\affiliation{Instituto de Aplica\c c\~ao Fernando Rodrigues da Silveira, 
Universidade do Estado do Rio de Janeiro, 20261-232 Rio de Janeiro, Brazil}
\author{C.A. Linhares}
\affiliation{Instituto de F\'isica, Universidade do Estado do Rio de Janeiro, 20559-900 Rio de
Janeiro, Brazil}
\email{linharescesar@gmail.com}
\author{A.P.C. Malbouisson}
\email{adolfo@cbpf.br}
\affiliation{Centro Brasileiro de Pesquisas F\'{i}sicas/MCTI, 22290-180 Rio
de Janeiro, Brazil.}
\begin{abstract}
We investigate the Casimir effect in the context of a nontrivial topology by
means of a generalized Matsubara formalism. This is performed in the context
of a scalar field in $D$ Euclidean spatial dimensions with $d$ compactified
dimensions. The procedure gives us the advantage of considering
simultaneously spatial constraints and thermal effects. In this sense, the
Casimir pressure in a heated system between two infinite planes is obtained
and the results are compared with those found in the literature. %
\end{abstract}
\maketitle
%
%%%%%%%%%%%%%%%%%%%%%%%%%%%%%%%%%%%%%%%%%%%%%%%%%%%%%%%%%%%%%%%%%%%%%%%%%%%%%%%%%%%%%%%%
\section{Introduction}
\label{intro} 
%%%%%%%%%%%%%%%%%%%%%%%%%%%%%%%

The Casimir effect is a quantum phenomenon originally described as the
attraction of two conducting, neutral, macroscopic objects in vacuum,
induced by changes in the zero-point energy of the electromagnetic field 
\cite{Casimir-1948}. This is not an exclusive feature of electromagnetic
fields. It has been shown that any relativistic field under the influence of
external conditions is able to exhibit an analogous kind of phenomenon \cite%
{Farina-BJP-2007}. This quantum vacuum effect is strongly dependent on the
material properties of the medium where the macroscopic objects interact, on
the nature of the quantum field, and on the boundary conditions under
investigation. It has been related to many different physical systems
ranging from cosmology, condensed matter, atomic and molecular physics to
more recent developments in micro and nanoelectricmechanical devices as
discussed in the reviews found in Refs. \cite{Milonni-1994,
Actor-Fortschr-Phys-1995, Mostepanenko-Trunov-1997,
Bordag-et-al-PhysRep-2001, Milton-2001, Milton-JPA-2004,
Lamoreaux-RepProgPhys-2005, Rodrigues-Capasso-Johnson-NaturePhotonics-2011,
CasimirPhysics-2011}. It is a well-known fact that thermal fluctuations also
produce Casimir forces. The pioneering works were devoted to explain its
thermodynamical behavior \cite{Fierz-Helv-Phys-Acta-1960, Mehra-Physica-1967}%
. General theoretical works \cite{Brown-Maclay-Phys-Rev-1969,
Balian-Duplantier-AnnPhys-1977-1978, Dowker-Kennedy-JPhysA-1978,
Kennedy-Dowker-Critchley-AnnPhys-1978, Tadaki-Takagi-ProgTheorPhys-1986,
Plunien-Muller-Greiner-PhysRep-1986, Plunien-Muller-Greiner-PhysicaA-1987,
Robaschik-etal-AnnPhys-1987}, and controversial results in realistic
situations \cite{Bostrom-Sernelius-PRL-2000, Bezerra-et-al-PRA-2002,
Hoye-et-al-PRE-2003, Decca-et-al-Ann-of-Phys-2005, Brevik-et-al-NJP-2006,
Chen-et-al-PRB-2007, Klimchitskaya-et-al-RMP-2009,
Brevik-Hoye-Eur-J-Phys-2014} were also explored. The first observation of
the Casimir force was made by Sparnaay in 1956 \cite{Sparnaay-Physica-1957}.
A few decades later, a large number of precise experimental evidences of
Casimir physics was found \cite{Precise-Casimir-experiments}.

The analysis of quantum field theory problems on toroidal spaces has been
the focus of a large number of investigations due to its applications to a
variety of problems, namely, second-order phase transitions in
superconducting films, wires and grains \cite%
{APCM-et-al-MPLA-2005,Abreu-et-al-JMP-2005,APCM-et-al-JMP-2009}, finite-size
effects in the presence of magnetic fields, finite chemical potential in
first-order phase transitions \cite{Correa-et-al-PLA-2013}, and also the
Casimir effect \cite{Ford-PRD-1980, Kleinert-Zhuk-TMP-1996,
JCSilva-PRA-2002, Queiroz-et-al-Annals-of-Phys-2005-2006,
Ahmadi-et-al-PRD-2002, Tomazelli-et-al-IJTP-2006}. It is well-known that one
way to obtain thermal effects in quantum field theories is to consider the
Matsubara formalism, in which a fourth dimension (mathematically analogous
to an imaginary time) has a finite extension equal to the inverse of
temperature $\beta $, with a periodic boundary condition. The application of
this procedure also to spatial dimensions has been introduced by Birrell and
Ford \cite{Birrel-Ford-PRD-1980} in order to describe field theories in
spaces with finite geometries and has been generalized to what came to be
known as quantum field theories on toroidal topologies \cite%
{APCM-et-al-NPB-2002, Khanna-et-al-Annals-of-Physics-2009,
Khanna-et-al-Thermal-Book-2009, Khanna-et-al-Annals-of-Physics-2011,
Khanna-et-al-Phys-Rep-2014}. Thus, this procedure can also be called a 
\textit{generalized Matsubara formalism}. In general, this technique
basically consists in considering quantum fields as defined over spaces with
topologies of the type $\left( \mathbb{S}^{1}\right) ^{d}\times \mathbb{R}%
^{D-d}$, with $1\leq d\leq D$, where $D$ represents the total number of
Euclidean dimensions and $d$ the number of compactified ones through the
imposition of periodic boundary conditions on the fields along them. One of
these dimensions is compactified in a circumference of length $\beta $,
whereas each of the spatial ones ($i=1,\ldots ,d-1$) in a circumference of
length $L_{i}$ and can be interpreted as boundaries of the Euclidean space 
\cite{Khanna-et-al-Thermal-Book-2009}. This corresponds to impose periodic
(antiperiodic) boundary conditions for fields in $D$ Euclidean dimensions
with $d$ compactified dimensions.

In the present paper we revisit the Casimir effect in this context, within a
Euclidean framework, as an application of the generalized Matsubara
formalism. We investigate the pressure experienced by the boundary in a
compactified space when a scalar field is heated. The starting point is the
so-called ``local formulation'', introduced in \cite%
{Brown-Maclay-Phys-Rev-1969}, in which the pressure is associated with the
33 component of the energy-momentum tensor. Then, we follow the
zeta-function regularization method originally employed by Elizalde and
Romeo \cite{Elizalde-Romeo-JMP-1989} for the computation of the Casimir
energy. However, here it is derived from a general formalism of field
theories on toroidal spaces as in Ref. \cite{Khanna-et-al-Phys-Rep-2014},
which allows to apply the method for several simultaneously compactified
dimensions. This is the case, for instance, of thermal field theories with a
finite spatial extension, which needs the compactification of both the
imaginary-time dimension and a spatial one for a unified approach for heated
Casimir cavities.

We stress that in our computation with the toroidal formalism periodic
boundary conditions are implemented both in imaginary time (circumference of
length $\beta$) and the third spatial coordinate (circumference of length $L$%
), by construction. Moreover, as stated in \cite{Khanna-et-al-Phys-Rep-2014}%
, results for other boundary conditions may be obtained from the periodic
ones. For instance, the pressure for Dirichlet boundary conditions (much
studied in the literature) can be determined by putting $L=2a$ in the
expression from the toroidal computation, where $a$ is the distance
separating the parallel plates in Ref.~\cite{Brown-Maclay-Phys-Rev-1969}.

The paper is organized as follows. In section \ref{T_em_for_scalar_fields}
the Casimir pressure is linked to the vacuum expectation value of the
energy-momentum tensor for a scalar field in $D$ dimensions of the Euclidean
space. The point-splitting technique is used to write it in terms of the
free scalar propagator in Fourier space. In section \ref%
{HeatedcacCasimirEffect} a corresponding expression for the pressure is
obtained when one of the spatial dimensions is compactified with a finite
extension. The computation of the Casimir pressure follows from the
Elizalde-Romeo method which leads to a well-known result from the
literature. In section \ref{ThEf}, we compute the Casimir pressure in the
configuration of a compact spatial dimension now in the presence of a
thermal bath, which can also be compared with results found in the
literature obtained from other techniques. In section \ref{finalremark} we
present our final comments. Throughout this paper, we consider $\hbar
=c=k_{B}=1$.

%%%%%%%%%%%%%%%%%%%%%%%%%%%%%%%
\section{Energy-momentum tensor for scalar fields}
\label{T_em_for_scalar_fields} 
%%%%%%%%%%%%%%%%%%%%%%%%%%%%%%%

We start by writing the Euclidean Lagrangian of the free scalar field in a $%
D $-dimensional space, 
\begin{equation}
\mathcal{L}_{E}=\frac{1}{2}\left( \partial _{\mu }\phi \right) ^{2}+\frac{1}{%
2}m^{2}\phi ^{2},  \label{eq:lagrangian}
\end{equation}%
where $m$ is the mass of the quanta of the scalar field. With the help of
the point-splitting technique, the vacuum expectation value of the canonical
energy-momentum tensor $T_{\mu \nu }$ can be written as \cite%
{Khanna-et-al-Phys-Rep-2014}, 
\begin{eqnarray}
\mathcal{T}_{\mu \nu } &=&\left\langle 0\left\vert T_{\mu \nu }\right\vert
0\right\rangle  \notag \\
&=&\lim_{x^{\prime }\rightarrow x}\!\mathcal{O}_{\mu \nu }\left( x,x^{\prime
}\right) \!\left\langle 0\left\vert {T}\phi \left( x\right) \phi \left(
x^{\prime }\right) \right\vert 0\right\rangle \!,
\label{eq:energymomentumtensor}
\end{eqnarray}%
where ${T}$ denotes the time-ordered product of field operators and $%
\mathcal{O}_{\mu \nu }\left( x,x^{\prime }\right) $ is a differential
operator given by \cite{Khanna-et-al-Phys-Rep-2014} 
\begin{equation}
\mathcal{O}_{\mu \nu }\left( x,x^{\prime }\right) =\partial _{\mu }\partial
_{\nu }^{\prime }-\frac{1}{2}\delta _{\mu \nu }\left[ \partial _{\sigma
}\partial _{\sigma }^{\prime }+m^{2}\right] ,  \label{eq:smart_operator}
\end{equation}%
where $\partial _{\mu }$ and $\partial _{\mu }^{\prime }$ are derivatives
acting on $x^{\mu }$ and $x^{\prime \mu }$, respectively, and $\delta _{\mu
\nu }$ represents the components of the metric tensor of the Euclidean space
(Kronecker delta). Defining the Euclidean Green function of the scalar field
as %
$G\left( x-x^{\prime }\right) =i\left\langle 0\left\vert {T}\left\{ \phi
\left( x\right) \phi \left( x^{\prime }\right) \right\} \right\vert
0\right\rangle $, we obtain 
\begin{equation}
\mathcal{T}_{\mu \nu }=\lim_{x^{\prime }\rightarrow x}\mathcal{O}_{\mu \nu
}\left( x,x^{\prime }\right) \left[ G\left( x-x^{\prime }\right) \right] .
\label{eq:energy_momentum_tensor_Euclidean_Green_function}
\end{equation}%
Considering the Fourier integral of the Euclidean Green function in momentum
space, 
\begin{equation}
G\left( x-x^{\prime }\right) =\int_{-\infty }^{\infty }\frac{d^{D}k}{\left(
2\pi \right) ^{D}}\frac{1}{k^{2}+m^{2}}e^{ik\cdot \!\left( x-x^{\prime
}\right) },  \label{eq:fourierGreenfunction}
\end{equation}%
where $k$ and $x$ are $D$-dimensional vectors, we are able to rewrite the
energy-momentum tensor v.e.v. of Eq.~(\ref%
{eq:energy_momentum_tensor_Euclidean_Green_function}) in the following
manner: 
\begin{equation}
\mathcal{T}{}_{\mu \nu }=\int_{-\infty }^{\infty }\frac{d^{D}k}{\left( 2\pi
\right) ^{D}}\left[ \frac{k_{\mu }k_{\nu }}{k^{2}+m^{2}}-\frac{1}{2}\delta
_{\mu \nu }\right] .  \label{eq:standard_tensor}
\end{equation}

%%%%%%%%%%%%%%%%%%%%%%%%%%%%%%%
\section{Casimir pressure in a compactified space}
\label{HeatedcacCasimirEffect} 
%%%%%%%%%%%%%%%%%%%%%%%%%%%%%%%

In this section, we investigate the Casimir pressure for the particular case
of just one compactified spatial dimension ($d=1$), along the lines of Ref.~%
\cite{Elizalde-Romeo-JMP-1989}. It is sufficient to consider the 33
component of the energy-momentum tensor to obtain the Casimir pressure
resulting from a topological constraint imposed by periodic boundary
conditions on the field at the parallel plates (taken as infinite planes)
separated by a fixed distance $L$ in the $x_{3}$-direction.

From Eq.~(\ref{eq:standard_tensor}), it is straightforward to write the bulk
expression 
\begin{equation}
\mathcal{T}_{33}=\frac{1}{2}\int_{-\infty }^{\infty }\frac{d^{D}k}{\left(
2\pi \right) ^{D}}\left[ \frac{k_{3}^{2}-\left( k_{\bot }^{2}+m^{2}\right) }{%
k_{3}^{2}+k_{\bot }^{2}+m^{2}}\right] ,
\label{eq:previous_pressure_expression}
\end{equation}%
where $k^{2}=k_{3}^{2}+k_{\bot }^{2}$, and $k_{\bot }$ refers to the ($D-1$%
)-dimensional vector orthogonal to the 3-direction in Fourier space.

Let us call $\mathcal{T}_{33}^{\mathit{c}}$ the response of vacuum
fluctuations in the object that plays the role of a topological constraint.
We perform this by means of the compactification of just one spatial
dimension. In order to obtain the Casimir pressure that acts on the boundary
of the compactified space, we shall use the generalized Matsubara procedure,
which is the original contribution of the present manuscript. Basically, in
the general case, the technique consists in the replacement of integrals in
momentum space by sums, namely, 
\begin{equation*}
\int \frac{dk_{j}}{2\pi }\rightarrow \frac{1}{L_{j}}\sum_{n_{j}=-\infty
}^{+\infty }
\end{equation*}%
where the index $j$ assumes the values $j=1,2,\ldots ,D-1$, and the momentum
coordinate $k_{j}$ exhibits discrete values, 
\begin{equation*}
k_{j}=k_{n_{j}}=\frac{2\pi n_{j}}{L_{j}},
\end{equation*}%
and $L_{j}$ refer to the finite extension of each of the $j$ spatial
dimensions. For practical purposes, let us compactify just the $x_{3}$%
-component of the vector $x$. With these ideas in mind, the generalized
Matsubara formalism enables us to substitute the bulk expression of Eq.~(\ref%
{eq:previous_pressure_expression}) %$\mathcal{T}_{33}^{\mathit{c}}$
by the following one: 
\begin{equation}
\mathcal{T}_{33}^{\mathit{c}}=\!\frac{1}{2L}\!\sum_{n=-\infty }^{+\infty
}\!\int_{-\infty }^{\infty }\frac{d^{D-1}k_{\bot }}{\left( 2\pi \right)
^{D-1}}\!\left[ \frac{k_{n}^{2}-\left( k_{\bot }^{2}+m^{2}\right) }{%
k_{n}^{2}+k_{\bot }^{2}+m^{2}}\right] \!.\;\;\;
\label{eq:tensor_matsubarizado_espacialmente}
\end{equation}%
Using the well-known results provided by dimensional regularization, 
\begin{eqnarray}
\int_{-\infty}^{\infty}\frac{d^{D}k}{\left(2\pi\right)^{D}} \frac{1}{\left[%
k^{2}+b^{2}\right]^{s}} & = & \frac{1}{\left(4\pi\right)^{\frac{D}{2}}} 
\frac{\Gamma\left(s-\frac{D}{2}\right)} {\Gamma\left(s\right)} 
\nonumber \\ \nonumber \\ & &
\times \left(\frac{1}{b^{2}}\right)^{s-\frac{D}{2}},
\label{eq:firstmultidimensionalintegral}
\end{eqnarray}
\begin{eqnarray}
\int_{-\infty}^{\infty}\frac{d^{D}k}{\left(2\pi\right)^{D}} \frac{k^{2}}{
\left[k^{2}+b^{2}\right]^{s}} & = & \frac{D}{2} \frac{1}{\left(4\pi\right)^{
\frac{D}{2}}} \frac{\Gamma\left(s-\frac{D}{2}-1\right)} {\Gamma\left(s\right)} 
\nonumber \\ \nonumber \\ & & 
\times \left(\frac{1}{b^{2}}\right)^{s-\frac{D}{2}-1},
\label{eq:secondmultidimensionalintegral}
\end{eqnarray}
we obtain 
\begin{eqnarray}
\mathcal{T}_{33}^{\mathit{c}} &=&\left\{ f_{s}\left( \nu ,L\right)
\sum_{n=-\infty }^{+\infty }\left[ \frac{\left( an^{2}-c^{2}\right) \Gamma
\left( \nu \right) }{\left( an^{2}+c^{2}\right) ^{\nu }}\right. \right.  
\nonumber  \\ &&
\left. -\left. \left( s-\nu \right) \Gamma \left( \nu -1\right) \frac{1}{%
\left( an^{2}+c^{2}\right) ^{\nu -1}}\right] \right\} _{\!\!s=1}\!\!,  
\nonumber \\
&& 
\label{eq:tensor_pre_summation} 
\end{eqnarray}
where $a=L^{-2}$, $c=m/{2\pi }$, $\nu =s-\left( D-1\right) /2$, and $%
f_{s}\left( \nu ,L\right) $ a function given by 
\begin{equation}
f_{s}\left( \nu ,L\right) =\frac{1}{2L}\frac{1}{\left( 4\pi \right) ^{s-\nu
}\left( 2\pi \right) ^{2\left( \nu -1\right) }\Gamma \left( s\right) }.
\label{f_vacuum}
\end{equation}%
Adding and subtracting the term $c^{2}\Gamma \left( \nu \right) $ to the
numerator of the first term on the right-hand side of Eq.~(\ref%
{eq:tensor_pre_summation}), we obtain 
\begin{eqnarray}
\mathcal{T}_{33}^{\mathit{c}} &=&\left\{ f_{s}\left( \nu ,L\right) \left[
\left( 2\nu -s-1\right) \sum_{n=-\infty }^{+\infty }\frac{1}{\left(
an^{2}+c^{2}\right) ^{\nu -1}}\right. \right.   \nonumber
\label{eq:tensor_more_than_pre_summation} \\
&&\left. -\left. 2c^{2}\left( \nu -1\right) \sum_{n=-\infty }^{+\infty }%
\frac{1}{\left( an^{2}+c^{2}\right) ^{\nu }}\right] \right\} _{\!\!s=1}\!\!,
\nonumber \\
&&
\end{eqnarray}
where we have used that $\Gamma \left( \nu \right) =\left( \nu -1\right)
\Gamma \left( \nu -1\right) $. Recalling the general definition of the
inhomogeneous multidimensional Epstein--Hurwitz zeta function \cite%
{Elizalde-Romeo-JMP-1989, Kirsten-JMP-1994, Elizalde-etalBook-1994,
Elizalde-etalBook-1995}, 
\begin{eqnarray}
Z_{d}^{c^{2}}\!\left( \nu \;;a_{1},\ldots ,a_{d}\right)  & = &
\!\!\!\!\!\!
\sum_{n_{1},\ldots ,n_{d} = -\infty}^{+\infty}
\!\!\!\!\!\!
\left(a_{1}n_{1}^{2}+\ldots +a_{d}n_{d}^{2}+c^{2}\right) ^{-\nu },  
\nonumber \\
&&
\label{eq:multidimensional_Epstein_Hurwitz_zeta_function} 
\end{eqnarray}
in the particular case of one-dimensional compactification ($d=1$), it
simplifies to 
\begin{equation}
Z_{1}^{c^{2}}\!\left( \nu \;;a\right) =\sum_{n=-\infty }^{+\infty }\left(
an^{2}+c^{2}\right) ^{-\nu }.
\label{eq:onedimensional_Epstein_Hurwitz_zeta_function}
\end{equation}%
Substituting the previous expression into Eq.~(\ref%
{eq:tensor_more_than_pre_summation}), the pressure can then be rewritten as 
\begin{eqnarray}
\mathcal{T}_{33}^{\mathit{c}} &=&\left\{ f_{s}\left( \nu ,L\right) \left[
\left( 2\nu -s-1\right) Z_{1}^{c^{2}}\!\left( \nu -1;a\right) \right.
\right.   \notag \\
&&\left. -\left. 2c^{2}\left( \nu -1\right) Z_{1}^{c^{2}}\!\left( \nu
;a\right) \right] \right\} _{\!\!s=1}.
\label{eq:tensor_more_than_pre_summation2}
\end{eqnarray}%
Following Ref.~\cite{APCM-et-al-NPB-2002}, these zeta functions can be
evaluated on the whole complex plane by means of an analytic continuation
described in the following manner \cite{Elizalde-Romeo-JMP-1989,
Kirsten-JMP-1994, Elizalde-etalBook-1994, Elizalde-etalBook-1995}: 
\begin{eqnarray}
Z_{d}^{c^{2}} && \left(\nu;a_1,\ldots,a_d\right) = \frac{2 \pi^{\frac{d}{2}}}
{\sqrt{a_1\cdots a_d} \; \Gamma\left(\nu\right)} 
\left[ \frac{1}{2c^{2\nu-d}} \Gamma\left(\nu-\frac{d}{2}\right) \right. 
\nonumber \\ \nonumber \\ & & 
\left. + 2\sum_{j=1}^{d}\sum_{n_j=1}^{\infty} \left(\frac{\pi n_j}{c \sqrt{a_j}}%
\right)^{\nu - \frac{d}{2}} K_{\nu-\frac{d}{2}}\left(2 \pi c \frac{n_j}{%
\sqrt{a_j}}\right) + \cdots \right. 
\nonumber \\ \nonumber \\  & & 
\left. + 2^{d}
\sum_{n_1,\ldots,n_d=1}^{\infty} \left(\frac{\pi}{c} \sqrt{\frac{n_1^2}{a_1} +
\cdots + \frac{n_d^2}{a_d}} \right)^{\nu-\frac{d}{2}} \right. 
\nonumber \\ \nonumber \\  & &
\left. \times K_{\nu-\frac{d}{2}}\left(2 \pi c \sqrt{\frac{n_1^2}{a_1} +
\cdots + \frac{n_d^2}{a_d}}\right) \right]\!,\;\;
\label{eq:analitycal_continuation_of_MultiZeta}
\end{eqnarray}
where $K_{\nu }\left( z\right) $ denotes modified Bessel functions of the
second kind. For $d=1$, the analytical continuation can be reduced to 
\begin{eqnarray}  
Z_{1}^{c^{2}}\!\left(\nu;a\right) & = & \frac{2\pi^{\frac{1}{2}}}{\sqrt{a}%
\;\Gamma\left(\nu\right)} \left[\frac{1}{2c^{2\nu - 1}}\Gamma\left(\nu-\frac{%
1}{2}\right) \right. 
\nonumber \\ \nonumber \\ & & 
\left. + 2\sum_{n=1}^{\infty}\left(\frac{\pi
n }{c \sqrt{a}}\right)^{\nu - \frac{1}{2}} \!K_{\nu-\frac{1}{2}}\left(2 \pi
c \frac{n}{\sqrt{a}}\right)\right].  
\nonumber \\
\label{eq:analitycal_continuation_of_Zeta}
\end{eqnarray}
After some algebraic manipulations, we notice the presence of terms which
are independent of the variable $L$, and for this reason are considered
unphysical. Neglecting these terms, we can show that 
\begin{eqnarray}
\mathcal{T}_{33}^{\mathit{c}} & = & 
2 \left(\frac{m}{2\pi L}\right)^{\frac{D}{2}} 
\Biggl[ \left(1-D\right)\sum_{n=1}^{\infty}\left(\frac{1}{n}\right)^{%
\frac{D}{2}} \! K_{\frac{D}{2}}\left(mnL\right) 
\nonumber \\ \nonumber \\ & & 
- \; mL\sum_{n=1}^{\infty}\left(\frac{1}{n}\right)^{\frac{D}{2}-1} K_{\frac{D}{2}%
-1}\left(mnL\right) \Biggr].  
\label{eq:Casimirffect-compactifiedspace}
\end{eqnarray}
The formula above corresponds to a general expression for the Casimir
pressure exerted by the vacuum fluctuations that induces a topological
effect due to the presence of the compactified manifold of length $L$. The
result presented in Eq. (\ref{eq:Casimirffect-compactifiedspace}) is the
Casimir vacuum radiation pressure for a massive scalar field submitted to
periodic boundary conditions in $D$ dimensions and is in agreement with
Refs.~\cite{Milton-2001, Ambjorn-Wolfram-Annals-of-Phys-1983,
AguiarPinto-etal-BJP-2003}.

For a $4$-dimensional Euclidean space, we obtain ~\cite%
{AguiarPinto-etal-BJP-2003} 
\begin{eqnarray}
\mathcal{T}_{33}^{\mathit{c}} \left(L,m\right) & = & - \frac{m^2}{2\pi^2L^2} %
\Biggl[ 3\sum_{n=1}^{\infty}\frac{1}{n^{2}} \! K_{2}\left(mnL\right) \nonumber \\ \nonumber \\ %
& & + \; mL\sum_{n=1}^{\infty}\frac{1}{n} K_{1}\left(mnL\right) \Biggr].
\label{eq:Casimirffect-compactifiedspace-forDequals4}
\end{eqnarray}
From the following asymptotic formula of the Bessel function, 
\begin{equation}
K_{\nu}\left(z\right) \approx {2^{\nu-1}z^{-\nu}\Gamma\left( \nu \right)},
\label{eq:asymptotic-small-BesselK}
\end{equation}
evaluated for small values of its argument $\left( z \sim 0 \right) $ and $%
\mathcal{R}e\;(\nu)\;>\;0 $, we obtain the small-mass limit Casimir pressure
($mL \ll 1$) 
\begin{equation}
\mathcal{T}_{33}^{\mathit{c}} \left(L,0\right) = -\frac{\pi^{2}}{30L^{4}} \;,
\label{eq:CasimirEffectParticularCasePeriodicBC}
\end{equation}
where we have neglected terms of $\mathcal{O}\left(m^2\right)$. The vacuum
fluctuation Casimir force per unit area is a finite negative expression
which suggests that the radiation pressure contracts the compactified space
of circumference $L$.

\begin{figure}[h!]
\label{CasimirVacuoNormalizado}
\includegraphics[width=0.9\columnwidth]{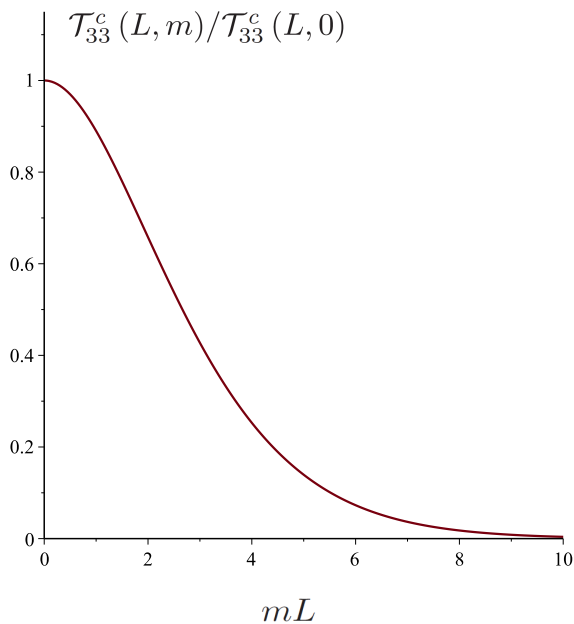}
\caption{Normalized Casimir pressure ${\mathcal{T}_{33}^{\mathit{c}}
\left(L,m\right)}/ {\mathcal{T}_{33}^{\mathit{c}} \left(L,0\right)}$ in
terms of the dimensionless factor $mL$. We clearly see the Casimir pressure
becomes a monotonically decreasing function for both: large mass and fixed $%
L $ or fixed mass $m$ and large values of the circumference of the
compactified manifold.}
\end{figure}

In Fig.~\ref{CasimirVacuoNormalizado}, we plot the ratio between the Casimir
pressure for massive scalar fields (Eq.~(\ref%
{eq:Casimirffect-compactifiedspace-forDequals4})) and for massless scalar
fields (Eq.~(\ref{eq:CasimirEffectParticularCasePeriodicBC})): ${\mathcal{T}%
_{33}^{\mathit{c}}\left( L,m\right) }/{\mathcal{T}_{33}^{\mathit{c}}\left(
L,0\right) }$. We notice in this figure that the normalized Casimir pressure
quickly becomes a monotonically decreasing function for absolute values of
the dimensionless parameter $mL$.

A no-less important comment we present to finalize this section is that the
corresponding negative Casimir pressure between two infinitely parallel
planes, when one imposes to the massless scalar field Dirichlet boundary
conditions, that is, $\phi (x_{3}=0)=\phi (x_{3}=L)=0$, is immediately
recovered when the plane separation distance $a$ is equal to the half
circumference length $L$ of the space dimension under compactification.

%%%%%%%%%%%%%%%%%%%%%%%%%%%%%%%%%%%%%%%%%
\section{Thermal effects}
\label{ThEf} 
%%%%%%%%%%%%%%%%%%%%%%%%%%%%%%%%%%%%%%%%%

In this section, thermal and boundary effects are taken care of
simultaneously through the generalized Matsubara prescription. We then
consider a $D$-dimensional space with a double compactification ($d=2$) of
the Euclidean space corresponding to a compactified spatial dimension with
length $L$ and a compactification of the imaginary-time dimension with
length $\beta$. In other words, we take the simultaneous compactification of
both the $x_{0}$ and $x_{3}$ coordinates of the vector $x$. %%%%%%%

By taking the same steps as in the previous sections, the stress tensor
component $\mathcal{T}_{33}^{\mathit{c}}$ given by Eq.~(\ref%
{eq:previous_pressure_expression}) of the system under investigation now
becomes 
\begin{eqnarray}
\mathcal{T}_{33}^{\mathit{c}} & = & \frac{1}{2 \beta L} \sum_{n_1,n_2=-%
\infty}^{+\infty} \int_{-\infty}^{\infty}\frac{d^{D-2}k_{\bot}}{%
\left(2\pi\right)^{D-2}} \nonumber \\ \nonumber \\ & & \times \left[\frac{%
k_{n_1}^{2}-k_{n_2}^{2}-\left(k_{\bot}^{2}+m^{2}\right)} {%
k_{n_1}^{2}+k_{n_2}^{2}+k_{\bot}^{2}+m^{2}}\right]\!.\;\;\;
\label{eq:tensor_matsubarizado_espacialmente2}
\end{eqnarray}
Using dimensional regularization, Eqs.~(\ref%
{eq:firstmultidimensionalintegral}) and (\ref%
{eq:secondmultidimensionalintegral}), the previous formula is rewritten as
follows: 
\begin{eqnarray}
\mathcal{T}_{33}^{\textit{c}}  & = &  
\Bigg\{ 
f_{s}\left(\nu,\beta,L\right)
\Bigg[
\sum_{n_1,n_2=-\infty}^{+\infty}
\frac{\left[a_1n_1^{2} - a_2n_2^{2}-c^{2}\right]\Gamma\left(\nu\right)}
{\left[a_1n_1^{2} + a_2n_2^{2}+c^{2}\right]^{\nu}}
\nonumber \\ \nonumber \\ &\;\;&
-
\sum_{n_1,n_2=-\infty}^{+\infty}
\frac{\left(s-\nu\right)\Gamma\left(\nu-1\right)}
{\left[a_1n_1^{2}+a_2n_2^{2}+c^{2}\right]^{\nu-1}}
\Bigg]
\Bigg\}_{\!\!s=1}\;\;,
\label{eq:tensor_pre_summation2}
\end{eqnarray}
where $a_{1}=L^{-2}$, $a_{2}=\beta ^{-2}$, $c=m/{2\pi }$, $\nu =s-\left(
D-2\right) /2$, and $f_{s}\left( \nu ,\beta ,L\right) $ is a function given
by 
\begin{equation}
f_{s}\left( \nu ,\beta ,L\right) =\frac{1}{2\beta L}\frac{1}{\left( 4\pi
\right) ^{s-\nu }\left( 2\pi \right) ^{2\left( \nu -1\right) }\Gamma \left(
s\right) }.  \label{DefinitionOfTheGeneralFunctionF}
\end{equation}%
Adding and subtracting the term $\left( a_{2}n_{2}^{2}+c^{2}\right) \Gamma
\left( \nu \right) $ in the numerator of the first term on the right-hand
side of Eq.~(\ref{eq:tensor_pre_summation2}), we obtain 
\begin{eqnarray}
\mathcal{T}_{33}^{\mathit{c}}  & = & 
\bigg\{ f_{s}\left( \nu ,\beta ,L\right)
\Gamma \left( \nu -1\right) \bigg[ \left( 2\nu -s-1\right) 
\nonumber \\ \nonumber \\ &&
\left.\left. 
\times Z_{2}^{c^{2}}\!\!\left( \nu -1;a_{1},a_{2}\right)
-2c^{2}\left( \nu
-1\right) Z_{2}^{c^{2}}\!\!\left( \nu ;a_{1},a_{2}\right) 
\right. \right.
\nonumber \\ \nonumber \\ &&
+2a_{2}\frac{\partial }{\partial a_{2}}Z_{2}^{c^{2}}\!\!\left( \nu
-1;a_{1},a_{2}\right) \bigg] \bigg\} _{s=1},
\label{eq:tensor_in_terms_of_Zeta}
\end{eqnarray}%
where we have used the definition of the two-dimensional Epstein--Hurwitz
zeta function, $Z_{2}^{c^{2}}\!\!\left( \nu ;a_{1},a_{2}\right) $, obtained
from Eq.~(\ref{eq:multidimensional_Epstein_Hurwitz_zeta_function}) for $d=2$%
. From Eq.~(\ref{eq:analitycal_continuation_of_MultiZeta}), we get for $d=2$ 
\begin{eqnarray}
Z_{2}^{c^{2}}\!\left(\nu;a_1,a_2\right)  &=& 
\frac{2\pi}{\sqrt{a_1 a_2} \; \Gamma\left(\nu\right)}
\left[
\frac{1}{2c^{2\left(\nu-1\right)}} \Gamma\left(\nu-1\right) 
\right. 
\nonumber \\ \nonumber \\ && 
\left.
+ 2\sum_{n_1=1}^{\infty} 
\left(\frac{\pi n_1}{c \sqrt{a_1}}\right)^{\nu - 1}K_{\nu-1}\left(2 \pi c \frac{n_1}{\sqrt{a_1}}\right) 
\right. 
\nonumber \\ \nonumber \\ && 
\left.
+ 2\sum_{n_2=1}^{\infty} 
\left(\frac{\pi n_2}{c \sqrt{a_2}}\right)^{\nu - 1}K_{\nu-1}\left(2 \pi c \frac{n_2}{\sqrt{a_2}}\right) 
\right. 
\nonumber \\ \nonumber \\ && 
\left.
+ 2^{2} \sum_{n_1,n_2=1}^{\infty}
\left(\frac{\pi}{c} \sqrt{\frac{n_1^2}{a_1} + \frac{n_2^2}{a_2}} \right)^{\nu-1}
\right. 
\nonumber \\ \nonumber \\ && 
\left. \times K_{\nu-1} \left( 2{\pi}{c} \sqrt{\frac{n_1^2}{a_1} + \frac{n_2^2}{a_2}} \right)
\right].
\label{eq:analitycal_continuation_of_BiZeta}
\end{eqnarray}
Substituting Eq.~(\ref{eq:analitycal_continuation_of_BiZeta}) in Eq.~(\ref%
{eq:tensor_in_terms_of_Zeta}), splitting $\mathcal{T}_{33}^{\mathit{c}}$
into three terms, $\mathcal{T}_{33}^{\mathit{c}}=\mathcal{T}_{\text{$n_{1}$}%
}^{\mathit{c}}+\mathcal{T}_{\text{$n_{2}$}}^{\mathit{c}}+\mathcal{T}_{\text{$%
n_{1}n_{2}$}}^{\mathit{c}},$ after removing removing nonphysical terms, we
have 
\begin{eqnarray} 
\mathcal{T}_{\text{$n_1$}}^{\textit{c}} & = & 
\frac{4\pi}{\sqrt{a_1 a_2}}
f_{s}\left(\nu,\beta,L\right)
\Bigg[
\left(2\nu - s - 2\right) \sum_{n_1=1}^{\infty} 
\left(\frac{\pi n_1}{c \sqrt{a_1}}\right)^{\nu - 2}
\nonumber \\ \nonumber \\ && \times
K_{\nu - 2}\left(2 \pi c \frac{n_1}{\sqrt{a_1}}\right) - 2c^2 \sum_{n_1=1}^{\infty} 
\left(\frac{\pi n_1}{c \sqrt{a_1}}\right)^{\nu - 1}
\nonumber \\ \nonumber \\ && 
\times K_{\nu - 1}\left(2 \pi c \frac{n_1}{\sqrt{a_1}}\right) 
\Bigg]\Bigg|_{s=1},
\label{eq:vacuumctb}
\end{eqnarray} 
which corresponds to the contribution to the Casimir pressure due to vacuum
fluctuations only. Using the definition (\ref%
{DefinitionOfTheGeneralFunctionF}), for $a_{1}=L^{-2}$, $a_{2}=\beta ^{-2}$, 
$c=m/{2\pi }$, $\nu =s-\left( D-2\right) /2$, Eq.~(\ref%
{eq:Casimirffect-compactifiedspace-forDequals4}) shown in the previous
section is recovered.

Also, 
\begin{eqnarray}
\mathcal{T}_{\text{$n_{2}$}}^{\mathit{c}} & = & 
\frac{4\pi }{\sqrt{a_{1}a_{2}}}f_{s}(\nu ,\beta ,L)
\Bigg\{ 
\left( 2\nu -s-2\right) 
\sum_{n_{2}=1}^{\infty }\left( \frac{\pi n_{2}}{c\sqrt{a_{2}}}\right) ^{\nu-2}
\nonumber \\ \nonumber \\ && 
\times 
K_{\nu -2}\left( 2\pi c\frac{n_{2}}{\sqrt{a_{2}}}\right) 
-2c^{2}\sum_{n_{2}=1}^{\infty }\left( \frac{\pi n_{2}}{c\sqrt{a_{2}}}\right)^{\nu -1}
\nonumber \\ \nonumber \\ && 
\times
K_{\nu -1}\left( 2\pi c\frac{n_{2}}{\sqrt{a_{2}}}\right) 
+ 2 a_{2} \frac{\partial }{\partial a_{2}}\sum_{n_{2}=1}^{\infty}
\Bigg[
\left( \frac{\pi n_{2}}{c\sqrt{a_{2}}}\right) ^{\nu -2}
\nonumber \\ \nonumber \\ && 
\times
K_{\nu -2}\left( 2\pi c\frac{n_{2}}{\sqrt{a_{2}}}\right) 
\Bigg]
\Bigg\}\Bigg|_{s=1}, 
\label{eq:thermalctb} 
\end{eqnarray}
yields
\begin{equation}
\mathcal{T}_{\text{$n_{2}$}}^{\mathit{c}} \left( \beta , m\right) =  
2\left( \frac{m}{2\pi \beta }\right) ^{\frac{D}{2}}
\!
\sum_{n_{2}=1}^{\infty }
\!
\left( \frac{1}{n_{2}}\right) ^{\frac{D}{2}}
\!\!K_{\frac{D}{2}}\left( m\beta n_{2}\right),
\label{eq:PureThermalCasimireffectformula}
\end{equation}
which is the Casimir force formula due exclusively to the thermal
fluctuations. The final form of Eq.~(\ref{eq:PureThermalCasimireffectformula}) 
was obtained by means of the useful recurrence formula for Bessel
functions, 
\begin{equation}
K_{\alpha - 1}\left( z \right) - K_{\alpha + 1}\left( z \right) = - \frac{%
2\alpha}{z} K_{\alpha}\left( z \right).  \label{eq:recurrenceBesselKformula}
\end{equation}
For $D=4$, we find 
\begin{equation}
\mathcal{T}_{\text{$n_{2}$}}^{\mathit{c}} \left( \beta , m\right) = \left( 
\frac{m^2}{2\pi^2 \beta^2 }\right) \sum_{n_{2}=1}^{\infty }\left( \frac{1}{%
n_{2}}\right) ^2 \!K_{2}\left( m\beta n_{2}\right).
\label{eq:PureThermalCasimireffectformulaOnceAgain}
\end{equation}
Using Eq.~(\ref{eq:asymptotic-small-BesselK}), we obtain the small-mass
limit purely thermal Casimir pressure ($m\beta \ll 1$) 
\begin{equation}
\mathcal{T}_{\text{$n_{2}$}}^{\mathit{c}} \left( \beta , 0\right) = \frac{%
\pi^2}{90 \beta^4},
\label{eq:PureThermalCasimireffectformulaOnceAgainandAgain}
\end{equation}
which is in accordance with the well-known Stefan-Bolztmann thermal
radiation pressure result. This is a finite positive force per unit area
which is more intense than vacuum radiation Casimir pressure for low values
of $\beta$ (high-temperature or classical limit).

If we plot the ratio between the thermal radiation pressure for the massive
scalar field (Eq.~(\ref{eq:PureThermalCasimireffectformulaOnceAgain})) and
the massless one (Eq.~(\ref%
{eq:PureThermalCasimireffectformulaOnceAgainandAgain})), as a function of
the dimensionless parameter $m\beta $, the normalized thermal Casimir force
per unit area $\mathcal{T}_{\text{$n_{2}$}}^{\mathit{c}} \left( \beta ,
m\right) / \mathcal{T}_{\text{$n_{2}$}}^{\mathit{c}} \left( \beta , 0\right) 
$ presents the typical monotonically decreasing shape for increasing values
of the parameter $m\beta $, in a qualitatively similar manner as Fig.~\ref%
{CasimirVacuoNormalizado}.
\begin{eqnarray}
\mathcal{T}_{\text{\ensuremath{n_{1}n_{2}}}}^{c} & = & 
\frac{8\pi}{\sqrt{a_1 a_2}}
f_{s}\left(\nu,\beta,L\right)
\Bigg\{ 
\left(2\nu - s - 2\right) 
\nonumber \\ \nonumber \\  & & \times
\sum_{n_1,n_2 = 1}^{\infty} 
\left(\frac{\pi}{c} \sqrt{\frac{n_1^2}{a_1} + \frac{n_2^2}{a_2}} \right)^{\nu-2}
\nonumber \\ \nonumber \\  & & \times
K_{\nu-2}\left(2 \pi c \sqrt{\frac{n_1^2}{a_1} + \frac{n_2^2}{a_2}}\right)
\nonumber \\ \nonumber \\  & & 
- 2c^2 \sum_{n_1,n_2 = 1}^{\infty} 
\left(\frac{\pi}{c} \sqrt{\frac{n_1^2}{a_1} + \frac{n_2^2}{a_2}} \right)^{\nu-1}
\nonumber \\ \nonumber \\  & & \times
K_{\nu-1}\left(2 \pi c \sqrt{\frac{n_1^2}{a_1} + \frac{n_2^2}{a_2}}\right)
\nonumber \\ \nonumber \\  & & 
+ 2a_2 \frac{\partial}{\partial a_2} \sum_{n_1,n_2 = 1}^{\infty}
\Bigg[
\left(\frac{\pi}{c} \sqrt{\frac{n_1^2}{a_1} + \frac{n_2^2}{a_2}} \right)^{\nu-2}
\nonumber \\ \nonumber \\  & & \times
K_{\nu-2}\left(2 \pi c \sqrt{\frac{n_1^2}{a_1} + \frac{n_2^2}{a_2}}\right)
\Bigg]\Bigg\}
\Bigg|_{s=1}, 
\label{eq:vacuumthermalcorrections}
\end{eqnarray}
provides
\begin{eqnarray}
\mathcal{T}_{\text{\ensuremath{n_{1}n_{2}}}}^{c} \!\left( L, \beta , m\right) \! & = & 
\!4\left(\frac{m}{2\pi}\right)^{\frac{D}{2}}\!\!
\Bigg[
\sum_{n_{1},n_{2}=1}^{\infty} \!
\left(\frac{1}{\sqrt{n_{1}^{2}L^{2}+n_{2}^{2}\beta^{2}}}\right)^{\frac{D}{2}}
\nonumber \\ \nonumber \\ \nonumber \\ && \times 
\left( 
\frac{\left(1-D\right)n_{1}^{2}L^{2}+n_{2}^{2}\beta^{2}}
{n_{1}^{2}L^{2}+n_{2}^{2}\beta^{2}}
\right)
\nonumber \\ \nonumber \\ \nonumber \\ && \times
K_{\frac{D}{2}}\left(m\sqrt{n_{1}^{2}L^{2}+n_{2}^{2}\beta^{2}}\right)
\nonumber \\ \nonumber \\ && {-} 
m\sum_{n_{1},n_{2}=1}^{\infty} \!
n_{1}^{2}L^{2} \!
\left(\frac{1}{\sqrt{n_{1}^{2}L^{2}+n_{2}^{2}\beta^{2}}}\right)^{\frac{D}{2}+1}
\nonumber \\ \nonumber \\ && \times
K_{\frac{D}{2}-1}\left(m\sqrt{n_{1}^{2}L^{2}+n_{2}^{2}\beta^{2}}\right)
\Bigg], 
\label{eq:vacuumthermalctb}
\end{eqnarray}
the corrections to the Casimir pressure in a compact space in the presence
of a massive scalar field heated at temperature $1/\beta $. In order to
obtain the final form of the above expression, we have used the recurrence
formula given by Eq.~(\ref{eq:recurrenceBesselKformula}). Considering $D=4$,
we get 
\begin{eqnarray}
\mathcal{T}_{\text{\ensuremath{n_{1}n_{2}}}}^{c} \!\left( L, \beta , m\right) \! & = & -
\left(\frac{m}{\pi}\right)^{2}
\Bigg[
\sum_{n_{1},n_{2} = 1}^{\infty}
\frac{3n_{1}^{2}L^{2} - n_{2}^{2}\beta^{2}}
{\left( n_{1}^{2}L^{2} + n_{2}^{2}\beta^{2} \right)^2}
\nonumber \\ \nonumber \\ && \times
K_{2}\left(m\sqrt{n_{1}^{2}L^{2}+n_{2}^{2}\beta^{2}}\right)
\nonumber \\ \nonumber \\ && {+} 
m\sum_{n_{1},n_{2}=1}^{\infty}
\frac{n_{1}^{2}L^{2}}{\left(n_{1}^{2}L^{2}+n_{2}^{2}\beta^{2}\right)^{\frac{3}{2}}}
\nonumber \\ \nonumber \\ && \times
K_{1}\left(m\sqrt{n_{1}^{2}L^{2}+n_{2}^{2}\beta^{2}}\right)
\Bigg], 
\label{eq:vacuumthermalctb4dimensions}
\end{eqnarray}
which is valid for arbitrary values of $m$, $L$ and $\beta $. Using Eq.~(\ref%
{eq:asymptotic-small-BesselK}), we can show that in the small-mass case it
reduces to 
\begin{equation}
\mathcal{T}_{\text{$n_{1}n_{2}$}}^{c}\!\left( L,\beta ,0\right) \!=-\frac{2}{%
\pi ^{2}}\sum_{n_{1},n_{2}=1}^{\infty }\frac{3n_{1}^{2}L^{2}-n_{2}^{2}\beta
^{2}}{\left( n_{1}^{2}L^{2}+n_{2}^{2}\beta ^{2}\right) ^{3}},
\label{eq:vacuumthermalpressuresmallmasslesslimit4dimensions}
\end{equation}%
where we have disregarded terms of $\mathcal{O}\left( m^{2}\right) $. The
corresponding expression for Dirichlet boundary conditions can be obtained
by substituting $L=2a$.

To clarify our results, we can show that the small-mass limit given by Eq. (%
\ref{eq:vacuumthermalpressuresmallmasslesslimit4dimensions}) can be written
as 
\begin{equation}
\mathcal{T}_{\text{$n_{1}n_{2}$}}^{c}\!\left( L,\beta ,0\right) \!=\frac{1}{%
L^{4}}\left[ 3f\left( \xi \right) +\xi s\left( \xi \right) \right] ,
\label{eq:CasimirThermalCorrectionForceperUnitAreaGeneralSymmetric}
\end{equation}%
where $\xi =L/\beta $ and 
\begin{equation}
f\left( \xi \right) =-\frac{1}{8\pi ^{2}}\sum_{n_{1},n_{2}=1}^{\infty }{%
\frac{(2\xi )^{4}}{\left[ \left( \xi n_{1}\right) ^{2}+\left( n_{2}\right)
^{2}\right] ^{2}}},  \label{eq:F_function}
\end{equation}%
\begin{equation}
s\left( \xi \right) =-f^{\prime }\left( \xi \right) =\frac{1}{\pi ^{2}}%
\sum_{n_{1},n_{2}=1}^{\infty }{\frac{(2\xi )^{3}n_{2}^{2}}{\left[ \left( \xi
n_{1}\right) ^{2}+\left( n_{2}\right) ^{2}\right] ^{3}}}.
\label{eq:S_function}
\end{equation}%
The function $f\left( \xi \right) $ obeys the inversion symmetry formula, 
\begin{equation}
f\left( \xi \right) =\xi ^{4}\;f\left( \frac{1}{\xi }\right) .
\label{eq:InversionSymmetryFormula}
\end{equation}%
This is an intriguing expression, known as temperature inversion symmetry,
that enables us to obtain the low and high-temperature limits after simple
algebraic manipulations, (see Refs. \cite%
{Brown-Maclay-Phys-Rev-1969,Tadaki-Takagi-ProgTheorPhys-1986,Gundersen-Ravndal-Ann-Phys-1988, Lutken-Ravndal-JPA-1988,Ravndal-Tollefsen-PRD-1989,Wotzasek-JPA-1990,Santos-etal-PRD-1999, Santos-Tort-Phys-Lett-B-2000,AguiarPinto-etal-PRD-2003}
for more details). Following ~\cite{Brown-Maclay-Phys-Rev-1969}, the
particular low-temperature limit ($\beta \gg 1$) can be more easily
performed after we compute the sum over index $n_{1}$ in Eq. (\ref%
{eq:F_function}),  
\begin{eqnarray}
f\left( \xi \right) & = & \frac{\xi^4}{\pi^2}\sum_{n_2=1}^{\infty}\frac{1}{n_2^4} -
\frac{\xi^3}{2\pi}\sum_{n_2=1}^{\infty}\frac{\coth\left( \pi n_2 / \xi\right)}{n_2^3}
\nonumber \\ \nonumber \\ & & - 
\frac{\xi^2}{2}\sum_{n_2=1}^{\infty}\frac{1}{n_2^2}\frac{1}{\sinh^2\left( \pi n_2 / \xi\right)} .
\label{eq:sum_over_n_1}
\end{eqnarray}
In the limit $\xi \ll 1$, the approximations 
\begin{eqnarray}
\coth \left( \pi n_{2}/\xi \right)  &\approx &1\;,  \label{eq:approxs1} \\
\sinh \left( \pi n_{2}/\xi \right)  &\approx &\frac{1}{2}e^{\pi n_{2}/\xi
}\;,  \label{eq:approxs2}
\end{eqnarray}%
are valid. Substituting Eqs.~(\ref{eq:approxs1}) and ~(\ref{eq:approxs2})
into Eq. (\ref{eq:sum_over_n_1}), and performing the sum over index $n_{2}$,
we find, for $\xi \ll 1$,%
\begin{eqnarray}
f\left( \xi \right) & = & \frac{\pi^2\xi^4}{90} -
\frac{\zeta\left(3\right) \xi^3}{2\pi} -2\xi^2 \left( 1 + \frac{\xi}{\pi}\right)e^{-2\pi /\xi}
\nonumber \\ \nonumber \\ & & + 
\mathcal{O}\left( e^{-4\pi /\xi} \right). 
\label{eq:f_final}
\end{eqnarray}
Inserting the above formula into Eq. (\ref%
{eq:CasimirThermalCorrectionForceperUnitAreaGeneralSymmetric}), we can show
that 
\begin{equation}
\mathcal{T}_{\text{$n_{1}n_{2}$}}^{c}\!\left( L,\beta ,0\right) \!=-\frac{%
\pi ^{2}}{90\beta ^{4}}+\frac{4\pi }{\beta L^{3}}\left( 1+\frac{L}{2\pi
\beta }\right) e^{-2\pi \beta /L}.  \label{eq:termo_misto_baixa_temperatura}
\end{equation}
In this sense, in the low-temperature limit ($L\ll \beta $), collecting all
the contributions, the final form of Casimir pressure  in the massless case
reads 
\begin{equation}
\mathcal{T}_{33}^{c}\!\left( L,\beta ,0\right) \!=-\frac{\pi ^{2}}{30L^{4}}+%
\frac{4\pi }{\beta L^{3}}e^{-2\pi \beta /L}.
\label{eq:Casimir_pressure_baixa_temperatura}
\end{equation}%
If we neglect the exponential factor, the Casimir pressure due exclusively
to the vacuum fluctuations is dominant in this regime.

The high-temperature limit is also easily found by means of the inversion
symmetry relation given by Eq. (\ref{eq:InversionSymmetryFormula}). Applying
this formula in Eq. (\ref{eq:f_final}), we get 
\begin{eqnarray}
f\left( \xi \right) & = & \frac{\pi^2}{90} -
\frac{\zeta\left(3\right) \xi}{2\pi} -2\xi^2 \left( 1 + \frac{1}{\pi\xi}\right)e^{-2\pi\xi}
\nonumber \\ \nonumber \\ & & + 
\mathcal{O}\left( e^{-4\pi\xi} \right). 
\label{eq:f_final_larger_xi}
\end{eqnarray}
Substituting Eq.~(\ref{eq:f_final_larger_xi}) into Eq. (\ref%
{eq:CasimirThermalCorrectionForceperUnitAreaGeneralSymmetric}), we find 
\begin{eqnarray}
\mathcal{T}_{\text{\ensuremath{n_{1}n_{2}}}}^{c}  \!\left( L, \beta , 0\right) \! & = &
\frac{\pi^2}{30 L^4} - \frac{\zeta\left(3\right)}{\pi\beta L^3} - \frac{1}{\beta L^3}
\nonumber \\ \nonumber \\ & & \times
\left(\frac{4 \pi L^2}{\beta^2}+\frac{6L}{\beta}+\frac{4}{\pi}\right)e^{-2\pi L / \beta} .
\label{eq:termo_misto_alta_temperatura}
\end{eqnarray}
Finally, in the high-temperature limit ($L\gg \beta $), computing all terms,
the final form of Casimir pressure is written as follows: 
\begin{eqnarray}
\mathcal{T}_{\text{33}}^{c}  \!\left( L, \beta , 0\right) \! & = &
\frac{\pi^2}{90 \beta^4} - \frac{\zeta\left(3\right)}{\pi\beta L^3} 
-\frac{1}{\beta L^3}
\nonumber \\ \nonumber \\ & & \times
\left(\frac{4 \pi L^2}{\beta^2}+\frac{6L}{\beta}+\frac{4}{\pi}\right)e^{-2\pi L / \beta} .
\label{eq:Casimir_pressure_alta_temperatura}
\end{eqnarray}
Notice that if we neglect the exponential factor, the Casimir pressure for
large temperature is given by the classical thermal radiation pressure $\pi
^{2}/\left( 90\beta ^{4}\right) $ plus a negative linear correction factor
proportional to $\beta ^{-1}$.

%%%%%%%%%%%%%%%%%%%%%%%%%%%%%%%
\section{Final remarks}
\label{finalremark} 
%%%%%%%%%%%%%%%%%%%%%%%%%%%%%%%

In the present work we investigate some aspects of the Casimir effect in the
context of nontrivial topologies. In particular, we revisited the Casimir
effect for a massive scalar field in a heated compact space by means of the
generalized Matsubara formalism. The usual attractive response of quantum
and thermal fluctuations are obtained and our results are in accordance with
those found in the literature. One may notice that all thermal contributions
to the Casimir pressure, given by $\mathcal{T}_{n_{2}}^{\mathit{c}}$ and $%
\mathcal{T}_{\text{$n_{1}n_{2}$}}^{\mathit{c}}$, vanish in the
zero-temperature ($\beta \rightarrow \infty $) limit, remaining the pure
dependence on the distance $L$ between plates, which has a well-known $L^{-4}
$ dependence in the small-$L$ limit for a four-dimensional space. Also, the
bulk limit $L\rightarrow \infty $ reduces all expressions in $D=4$ to the
Stefan--Boltzmann law $\beta ^{-4}$.

A rather peculiar aspect of the generalized Matsubara formalism is related
to the renormalization of the expressions. Usually, in the Casimir context,
the divergent terms are taken care of by subtraction of the bulk integral,
without compactifications (see \cite{Khanna-et-al-Phys-Rep-2014}). Here,
there is no need to do so, as was also remarked by Elizalde and Romeo \cite%
{Elizalde-Romeo-JMP-1989}. It is sufficient to obtain correct physical
expressions to renormalize by subtraction the divergent term of the
expansion of the Epstein--Hurwitz zeta functions $Z_{d}^{c^{2}}$, as it does
not depend on the physical parameters $L$ or $\beta $.

We also remark that the expression we obtain from the toroidal formalism,
which conveys periodic boundary conditions in the compactified dimensions,
lead to corresponding ones for the Dirichlet conditions, by substituting $%
L=2a$. The $D=4$, small-$L$ limit of the Casimir pressure in the nonthermal
case, given by Eq.~(\ref{eq:Casimirffect-compactifiedspace}) becomes $%
\mathcal{T}_{33}=-\pi ^{2}/480a^{4}$ in the Dirichlet case for a quantum
scalar field. For an electromagnetic field, we have then twice that value, $%
\mathcal{T}_{33}=-\pi ^{2}/240a^{4}$, due to its two degrees of freedom.
These are compatible with the original Casimir results.

%%%%%%%%%%%%%%%%%%%%%%%%%%%%%%%
\section{Acknowledgments}
\label{ack} 
%%%%%%%%%%%%%%%%%%%%%%%%%%%%%%%

This work was partially supported by the Brazilian agencies CNPq and FAPERJ. 

%%%%%%%%%%%%%%%%%%%%%%%%%%%%%%%
%

%%%%%%%%%%%%%%%%%%%%%%%%%%%%%%%
\end{document}